\documentstyle[12pt]{article}
\begin{document}
\begin{titlepage}

\title { CURRENT MAGNIFICATION AND CIRCULATING CURRENTS IN
MESOSCOPIC RINGS}
\author{A. M. Jayannavar$^1$, P. Singha Deo and T. P. Pareek
\\%footnotemark[2] \\
Institute of Physics, Sachivalaya Marg, Bhubaneswar-751005, INDIA.}

\footnotetext[1]{e-mail:jayan@iopb.ernet.in}
\maketitle

\thispagestyle{empty}

\begin{abstract}
We show that several novel effects
related to persistent currents can arise
in open systems, which have no analogue in closed
or isolated systems. We have considered a system of a
metallic ring coupled to two electron reservoirs.
We show that in the presence of a transport
current, persistent currents can flow in a ring
even in the absence of magnetic field.
This is related to the current
magnification effect in the ring.
In the presence of magnetic field
we show that the amplitude of persistent currents
is sensitive to the direction of
current flow from one reservoir to another.
Finally we briefly discuss the persistent currents
arising due to two nonclassical effects namely,
Aharonov-Bohm effect and quantum tunneling.

\end{abstract}

PACS NO :72.10.-d, 05.60.+w, 72.10.Bg, 67.57.Hi

\end{titlepage}

\eject

\newpage
\hspace {0.5in}
{\bf I. Introduction}

It is well known that Aharonov-Bohm(AB)
effect reveals, in solid state physics, as a
characteristic dependence of physical properties
of mesoscopic systems on magnetic flux.
An apt example being the AB oscillations in magneto-resistance
of small conducting rings[1]. In recent years a
lot of interest has been
generated in the persistent circulating currents
induced in a mesoscopic ring by an AB flux $\phi$[2].
These are manifestations of novel quantum effects in
submicron systems beyond the atomic realm.
B$\ddot u$ttiker et al proposed, a decade ago, that
an ideal one dimensional metallic ring (containing elastic
scatterers) of mesoscopic size (whose dimensions are less
than the inelastic mean free path or phase breaking length) can
support an equilibrium circulating current in response to
magnetic flux[3]. The coherent wave function (even in the presence of
elastic scatterers) extending over the whole
circumference of the loop leads to a response with
period equal to a unit of magnetic flux $\phi_{0}$=hc/e.
There are now three experiments confirming the existence of
persistent currents in mesoscopic rings[4-6]. The experiments
were performed on ensemble of many mesoscopic rings[4] as well as on
single (or several) rings[5,6]. Theories based on
non-interacting electron picture predict magnitude of currents
much less than the experimentally observed
ones. Interest in this area has increased as a result
of this discrepancy of up to two orders
of magnitude between theory and experiment[1]. Results
of the later experiment on very weakly disordered ring
agree with the theory in the ballistic regime[6].
Experimental results for
disordered systems (or in the diffusive regime) have not yet
been satisfactorily explained.

Application of magnetic field destroys time-reversal symmetry
and as a consequence, the degeneracy of states carrying current
clockwise or anticlockwise is lifted. Depending on the Fermi level
uncompensated current flows in either of the
directions (diamagnetic or paramagnetic). For a perfect one dimensional
ring of length L, the magnetic flux $\phi$ in the loop modifies
the periodic boundary condition into
$\psi(x+L)=\psi(x) e^{(i2\pi\phi/\phi_{0})}$.
This condition implies that the energy levels
and hence all physical observables are periodic functions
of the flux with a period $\phi_{0}$. The persistent current is given
by the flux derivative of the total free energy of the ring.
For an ideal isolated ring persistent current at zero
temperature exhibits periodic saw tooth behavior as
a function of $\phi$. For even number of
electrons N the jump discontinuities occur
from the value -(2e$v_{f}$/L) to (2e$v_{f}$/L) at
$\phi=0,\pm\phi_{0},\pm2\phi_{0},..$ etc.
and at $\phi=\pm\phi_{0}/2,\pm3\phi_{0}/2$, etc. for odd N. Here
$v_{f}$ is the Fermi energy. The typical magnitude of the
persistent currents
at T=0 for L between 1-3 $\mu$m and for Fermi wave vector $k_{f}$
between $10^{10}m^{-1}$ (metallic ring) and $10^8m^{-1}$
(semiconducting ring)
varies between 1 and 5 nA. We would also like to
emphasize that since the magnetic field tunes
the boundary condition, the persistent currents can be
thought to arise due to the consequence of the sensitivity of the
eigenstates to the boundary conditions. Persistent currents are purely
mesoscopic effects in the sense that they are strongly suppressed
when the ring size exceeds the characteristic
dephasing length of the electrons or the inelastic mean free
path. The quantum persistent current
is a sample specific property and for a  given flux,
apart from N dependence it exhibits sensitive
dependence on the microscopic configuration of
disorder and hence non-self-averaging fluctuation effects.
Studies have been extended to include multichannel rings,
spin-orbit coupling, disorder, electron-electron interaction
effects, etc[2,7-12]. The problem of persistent currents has also
facilitated the study of some fundamental problems of
statistical mechanics, most notably the questions concerning
the role of statistical ensemble. The disordered average
current has been found to be vanishingly small for
moderate disorder, when grand canonical ensemble has been used,
while it is of finite amplitude within the framework of
canonical ensemble[2].

Theoretical treatments to date have been mostly
concentrated on isolated rings. Persistent currents occur not
only in isolated rings but also in the rings connected
via leads to electron reservoirs, namely open systems[13-15].
In a recent experiment Mailly et al have measured
the persistent currents in both closed and open rings[6].
The reservoir acts as a source and sink for electrons
and is characterized by a well defined chemical potential
$\mu$. There is no phase relationship between the absorbed
and the emitted
electrons. Thus the reservoirs act as a source of
energy dissipation as well as inelastic scatterer.
All the scattering processes in the leads including the loop
are assumed to be elastic. Inelastic processes occur
only in the reservoirs, and hence there is a complete spatial
separation between elastic and inelastic processes. Weak
inelastic processes do not destroy the periodic behavior
of persistent currents as a function of magnetic flux $\phi$.

In our present work we show that several novel effects
related to circulating currents arise in open systems.
All these effects have no analogue in closed or
isolated systems. We have considered a system of metallic loop
of circumference L coupled to two electron reservoirs,
characterized by chemical potentials $\mu_{1}$ and $\mu_{2}$
connected via ideal leads as shown in fig. (1). For the sake of simplicity
we have considered 1-D free electron networks.
We have introduced a $\delta$-function impurity
of strength V at a distance $l_{3}$ from the junction $J_{2}$,
to break the spatial symmetry in the problem.
The lengths of the upper and lower arms of the loop
are $l_{1}$ and $l_{2}$, respectively. We have set
the units $\hbar$, m to unity and all the lengths are scaled
with respect to the length L of the circumference of the loop (L=
$l_1+l_2$).
We first show that circulating currents in a ring can arise
even in the absence of magnetic field, but in the presence
of a transport current. This is related to current magnification
in the loop and is of purely quantum mechanical origin.
Next, we show that in the open systems the magnitude of persistent currents
is sensitive to the direction of current flow.
Finally we discuss briefly the properties of persistent currents
arising simultaneously due to
two non-classical effects, namely, AB effect
and quantum tunneling.

{\bf II. Circulating currents in the absence of magnetic field}

In this section we set the strength of the impurity to zero just
for the sake of simplicity. The effect discussed here manifests
itself even in the presence of impurities. To obtain
transport current we must have $\mu_{1}\ne\mu_{2}$
(non-equilibrium situation).
The transport current
will be directed from left to right or from right to left
depending on whether
$\mu_{1}>\mu_{2}$ or $\mu_{2}>\mu_{1}$.
We will show that in
this case a circulating current is induced in the ring by
incident carriers. Existence of such currents was first
discussed by B$\ddot u$ttiker[14]. However, our analysis is
qualitatively different from the earlier studies. The
current injected by the reservoir into the lead around the small
energy interval dE is given by $dI_{in}$=ev(dn/dE)f(E)dE, where
v=$\hbar k/m$ is the velocity of the carriers at the energy E,
dn/dE=1/(2$\pi \hbar$v) is the density of states in the perfect
wire and f(E) is the Fermi distribution. The total current flow
I in a small energy interval dE through the system is given by
the current injected into the leads by reservoirs multiplied by
the transmission probability T. This current splits into
$I_{1}$ and $I_{2}$ in the upper and lower arms such that
I=$I_{1}+I_{2}$ (current conservation). As the upper and lower arm
lengths are unequal, these two currents are different in
magnitude.
In our present quantum problem when one calculates the
currents ($I_{1},I_{2}$) in the two arms there
exists two distinct possibilities. The first possibility
being for a certain range of incident Fermi wave vectors
the current in the two arms $I_{1}$ and $I_{2}$ are individually
less than the total current I, such that
I=$I_{1}+I_{2}$. In such a situation both currents
in the two arms flow in the direction of applied electric
field. However, in
certain energy intervals, it turns out that the current in
one arm is larger than the total current I (magnification
property). This implies that to conserve the total current at the junctions
the current in the other arm
must be negative, i.e., the current should flow against the
applied external field induced by difference in the chemical
potentials. In such a situation one can interpret the
negative current flowing in one arm, continues to flow
as a circulating current or persistent current in the loop.
Thus the magnitude and direction of persistent current is
the same as that of the negative current.
Our procedure of assigning
persistent current, is exactly the same as the procedure
well known in classical LCR ac network analysis. When a
parallel resonant circuit (capacitance C connected in parallel
with combination of inductance L and resistance R) is driven by
external electromotive force (generator), circulating currents
arise in the circuit at a resonant frequency[16]. This
phenomenon is well known as current magnification.
It turns out that even in our quantum problem the circulating
currents arise near the antiresonances in the transmission
coefficient of the loop structure
coupled to leads.

We now consider the case where the current is injected from left
reservoir ($\mu_{1}>\mu_{2}$). At temperature zero the total
current flow around a small energy interval dE around E is
I=(e/$2\pi\hbar$)T, where T is the transmission coefficient
calculated at the energy E.
It is a straight forward exercise to set up a scattering
problem and calculate the transmission coefficient (T)
and the currents in the upper ($I_{1}$) and the lower ($I_{2}$)
arms. We closely follow our earlier method of quantum waveguide
transport on networks to calculate these quantities[15,17-19]. We
have imposed the Griffiths boundary conditions (conservation of current)
and single valuedness of the wavefunctions at the junctions.
The expressions are given by

\begin{equation}
I=(e/2\pi\hbar)T,
\end{equation}

\begin{equation}
T=(8   (2 - cos[2 k l_{1}] -
     cos[2 k l_{2}] + 4 sin[k l_{1}] sin[k l_{2}]))/\Omega,
\end{equation}

\begin{equation}
I_{1}=(e/2\pi\hbar)8(1 - cos[2 k l_{2}] +
2 sin[k l_{1}] sin[k l_{2}])/\Omega,
\end{equation}

\begin{equation}
I_{2}=(e/2\pi\hbar)8(1 - cos[2 k l_{1}]  +
2 sin[k l_{1}] sin[k l_{2}])/\Omega,
\end{equation}

\noindent where

 $$\Omega= (37 - 5   cos[2 k l_{1}] -
    32 cos[k l_{1}] cos[k l_{2}] - 5 cos[2 k l_{2}] +
 $$
\begin{equation}
5 cos[2 k l_{1}] cos[2 k l_{2}] + 48 sin[k l_{1}] sin[k l_{2}] -
    4  sin[2 k l_{1}] sin[2 k l_{2}]).
\end{equation}

In fig.(2) we have plotted the circulating currents (solid
curves) in the dimensionless units
($I_{c}\equiv 2\pi\hbar I_{c}/e$) in the small energy interval dE
around the Fermi energy as a function of dimensionless wave
vector kL. We have chosen $l_{1}/l_{2}$=5.0/3.0. In fig. (2) we
have also plotted the transmission coefficient T for the same
parameter values. We notice that the persistent current changes
sign as we cross the energy or the wave vector corresponding to
the first
antiresonance(or transmission zero) in the transmission
coefficient. It does not change sign as we cross the second
antiresonance. The first antiresonance is characterized by a
asymmetric zero-pole in the transmission amplitude (zero occur
at a value of kL=$(2\pi)$ and poles are given by kL=(6.25495-i
0.299976) and (6.46865-i 1.90045)). The proximity of the zero
and pole leads to sharp variations and asymmetry in the transmission
coefficient around the magnitude zero as a function of energy
(around the first antiresonance).
The second antiresonance is characterized by a zero
along with symmetrically placed two poles and transmission
coefficient is symmetric around the antiresonance. The zero is
at a value kL=(4$\pi$) and poles are given by kL=(12.4105-i
1.07584) and (12.7222-i 1.07584). Thus, we have shown that persistent
currents arise near the vicinity of the antiresonances and the
nature of persistent currents depend on the zero-pole structure
in the transmission amplitude. In general zero-pole structure in
the transmission coefficient is sensitive to the ratio
$l_{1}/l_{2}$ being commensurate or not. For incommensurate
ratio we mostly obtain asymmetric antiresonances. The magnitude
and the width of the persistent current peaks in the vicinity of
antiresonances depend on the strength of the imaginary part of
the poles. If the two poles have different imaginary parts, the
peak value of the persistent current will be higher for
the pole with smaller imaginary
part. For fixed value of Fermi
energy the persistent current changes sign as we change the
direction of the current flow. In equilibrium
($\mu_{1}=\mu_{2}$) we cannot have persistent currents in the
absence of magnetic field. If $\mu_{1}>\mu_{2}$, then at zero
temperature the total magnitude of persistent current is given
by $I_{T}=\int_{\mu_{1}}^{\mu_{2}}I_{c}dE$. The total magnetic
moment of a loop is proportional to $\oint I(l)dl$ ($l$ is taken
along the loop). Owing to the current
magnification property we expect that one should experimentally
observe enhanced magnetic moment near the antiresonances in the
transmission coefficient or the two port conductance.

{\bf III. Persistent current dependence on the direction of the
current }

Here we shall discuss a phenomenon that arises only when
the impurity strength $V\neq0$. The presence of the
impurity breaks the spatial symmetry of the system.
We also restrict to the case of $l_{1}=l_{2}$, to avoid the
additional contribution arising due to the difference in
transport current across upper and lower arms. If
$\mu_{1}>\mu_{2}$ the net current flows from left to right and
vice versa if $\mu_{1}<\mu_{2}$. For the case of
$\mu_{1}>\mu_{2}$ at zero temperature, the reservoir 1 injects
a steady flux of electrons in the interval $\mu_{1}$ and
$\mu_{2}$ and results in a current flowing in the right
direction. These electrons moving to the right are first
scattered at $J_{1}$, $J_{2}$ and then at impurity site I along
with multiple scattering. If $\mu_{1}<\mu_{2}$, (i.e., the case
of current flowing in the left direction), the injected
electrons from reservoir 2 are first scattered at I (the
impurity site) and then at
$J_{2}$ and $J_{1}$ along with multiple reflections.
As there is no spatial symmetry, for these two different cases
the electron wavefunction (scattering states) have a different
complex amplitude at $J_{1}$ and $J_{2}$. This amounts to
imposing different boundary conditions across a loop. As
mentioned earlier the persistent currents are sensitive to
boundary conditions and hence we obtain different magnitude for
the persistent currents depending on the direction of current
flow[15]. We
have obtained analytical expressions for the persistent
currents. However, here we present our results graphically.
In fig.(3) we have plotted persistent current in dimensionless
units $dj$/k as a function of the
flux $\alpha(=2\pi \phi /\phi_{0})$ for a fixed value of
kL=(7.0) and VL=(10.0).
The dashed and solid curves
represent magnitudes of the persistent currents flowing in the loop
$dj_{R}/k$ and $dj_{L}/k$, respectively, i.e., when the d.c current
flows in the right and left directions, respectively. One can
readily notice, the directional dependence of the
persistent currents. In fig. (4) we have plotted the persistent
currents $dj_{R}/k$ and $dj_{L}/k$ as a function of dimensionless
impurity potential VL, for a fixed value of kL=(7.0) and for $\alpha$=(0.7).
The magnitude $dj_L/k$ decreases monotonically to zero as
VL$\rightarrow \infty$. This is due to the fact that in
this limit electrons emitted by right reservoir do not enter the
loop and cannot contribute for the persistent currents. The
absolute magnitude of $dj_R/k$ saturates to a value in the same
limit. This corresponds to a situation where the loop is connected
to single reservoir 1 and the connection
truncated at the point I(the impurity
site). At zero
temperature the total contribution to the persistent current is
obtained by adding all the contributions to $dj_R/k$ and $dj_L/k$ from
levels with energies upto the chemical potentials. Thus for fixed
$\mid \mu_{1} - \mu_{2} \mid$ we get a different persistent current
depending on the direction of the current flow. In our above
discussion we have restricted to a simple case of $l_1=l_2$. As
mentioned earlier, when $l_{1} \ne l_{2}$, an additional
contribution to the persistent current arises due to the
difference in the transport current across the upper and lower
arms. In such a situation the total persistent current (or
associated magnetic moment) becomes asymmetric under the
reversal of magnetic field. The magnitude of the transport
currents in the two arms in the fermi energy range of current
amplification depends on the magnetic field. In our simple
geometry we can show that for particular value of $\alpha(=2 \pi
\phi/\phi_{0})=\pi/2,\, 3\pi/2,$ etc. the total persistent current
is antisymmetric with respect to the magnetic field, and for
these values of magnetic field transport currents in the two arms
do not exhibit the current magnification property.

{\bf IV. Persistent currents due to evanescent modes}

Let us imagine a geometry where a metallic loop is coupled to a single
electron reservoir via an ideal lead. In an ideal lead the
potential is assumed to be zero i.e., V=0. In the metallic loop the
potential throughout the circumference is V and
is positive. When injected electrons have their energies less
than V, these
electrons can tunnel
into the loop quantum mechanically and propagate inside the loop
as evanescent modes and give rise to a persistent current in the
presence of a magnetic field.
Such currents arise simultaneously due to two nonclassical
effects, namely, quantum tunneling and Aharonov Bohm effect[20].
Currents due to such evanescent modes are
to be found by analytical continuation and we have found out
analytical expressions for these persistent currents. In the limit
QL$>>$1, persistent currents in a small energy interval around E
are given by
$dj=f(k,Q)e^{-QL}\sin(\phi/\phi_{0})$, where f(k,Q) is a simple
function of k and Q. Here k is the wave vector for incident
electrons i.e., k=$\sqrt{2mE/\hbar^{2}}$ and
Q=$\sqrt{2m(E-V)/\hbar^{2}}$. As expected the persistent currents
are periodic in magnetic flux with period $\phi_{0}$. Owing to
the decaying nature of evanescent modes, the factor arising due to
the sensitivity of the wavefunctions to the boundary conditions
appears as $e^{-QL}$. Higher harmonics in magnetic flux (say
nth harmonic) also contribute to the persistent currents
with a multiplication factor $\sin(n\phi/\phi_{0})$. However,
these harmonics, are weighted by $e^{-nQL}$ because for these
harmonics to appear the electron has to traverse the loop n
times. So, these harmonics can be neglected in the limit QL$>>$1.
Unlike the behavior of persistent currents above the
barrier regime the currents due to evanescent modes do not
oscillate as a function of the Fermi energy as long as E$<$V. The total
persistent current is given by sum of contributions from the
electrons upto Fermi energy. Even though the current due to
individual evanescent modes is small the total sum can have an
observable amplitude. Especially in a real physical situation
one can have a ring with extremely narrow width connected to
the reservoir via an ideal wire with a much larger width. In this
situation the zero point quantum potential due to the transverse
confinement in the ring is much higher than the zero point energy
of the ideal wire. Electrons can occupy several subbands in the
connecting wire but still they have energies less than the zero
point energy of the ring. All these electrons in several subband
modes will propagate as evanescent modes in the ring, and in this
situation a higher contribution to the total persistent current
may arise.

In conclusion, we have shown that several novel effects related
to persistent currents can arise in open systems which have no
analogue in closed or isolated systems. All these new features
can be experimentally verified. Further work in the direction of
including disorder, and interaction effects is needed to put these
effects on firm foundation.
\vfill
\eject

\vfill
\eject
{\bf Figure captions}

Fig. 1. An open metallic loop connected to two electron
reservoirs.

Fig. 2. Plot of persistent current $I_c(=2\pi\hbar I_c/e)$
versus kL (solid curve) and transmission coefficient T versus kL
(dashed curve) for $l_1/l_2$=5.0/3.0 .

Fig. 3. Persistent current versus $\alpha=(2\pi\phi/\phi_0)$ for
a fixed value of kL=7.0 and VL=10.0. The dashed curve represents
$dj_L/k$ and the solid curve represents $dj_R/k$.

Fig. 4. Persistent current versus impurity potential VL for a
fixed value of $\alpha$=0.7 and kL=7.0. The dashed curve represents
$dj_L/k$ and the solid curve represents $dj_R/k$.

\vfill
\eject
\end{document}